\documentclass[aps,prd,notitlepage,showkeys,amssymb,amsmath,floatfix,nofootinbib,superscriptaddress]{revtex4-1}
\usepackage{epsfig}
\usepackage{bm}
\usepackage{amssymb}
\usepackage{amsmath}
\usepackage{color}
\usepackage{slashed}
\usepackage{amsthm}
\newtheoremstyle{user}
{}
{}
{\normalfont}
{}
{\bfseries}
{.}
{2ex}
{\thmname{#1}\thmnumber{ #2}\thmnote{ \textnormal{#3}}}
\theoremstyle{user}

\usepackage{graphicx,subfigure}
\usepackage[thinlines]{easytable}
\usepackage[colorlinks,linkcolor=blue,anchorcolor=black,citecolor=blue]{hyperref}
\usepackage{tikz}
\newcommand{\sumr}{\mathop{\to\!\!\!\!\!\!\!\!\sum}\limits}
\usetikzlibrary{calc,decorations.markings}
\newcommand\smallO{
  \mathchoice
    {{\scriptstyle\mathcal{O}}}
    {{\scriptstyle\mathcal{O}}}
    {{\scriptscriptstyle\mathcal{O}}}
    {\scalebox{.6}{$\scriptscriptstyle\mathcal{O}$}}
  }
\allowdisplaybreaks[4]

\begin{document}

\title{On convergence properties of GPD expansion \\through Mellin/conformal moments and orthogonal polynomials}

\author{Hao-Cheng Zhang}
\email{hczhang2003@gmail.com}
\affiliation{Taishan College, Shandong University, Jinan, Shandong, 250100, China}

\author{Xiangdong Ji}
\email{xji@umd.edu}
\affiliation{Maryland Center for Fundamental Physics, Department of Physics, University of Maryland, 4296 Stadium Dr., College Park, MD 20742, U.S.A.}
\date{\today}

\begin{abstract}
We examine convergence properties of reconstructing the generalized parton distributions (GPDs) through the universal moment parameterization (GUMP). We provide a heuristic explanation for the connection between the formal  summation/expansion and the Mellin-Barnes integral in the literature, and specify the exact convergence condition. We derive an asymptotic condition on the conformal moments of GPDs to satisfy the boundary condition at \(x=1\) and subsequently develop an approximate formula for GPDs when \(x>\xi\). Since experimental observables constraining GPDs can be expressed in terms of double or even triple summations involving their moments, scale evolution factors, and Wilson coefficients, etc., we propose a method to handle the ordering of the multiple summations and convert them into multiple Mellin-Barnes integrals via analytical continuations of integer summation indices. 

\end{abstract}

\maketitle

\section{Introduction}
Generalized parton distributions (GPDs) \cite{Muller:1994ses,Ji:1996ek,Ji:1998pc}, which depict the three-dimensional (3D) structure of the nucleon \cite{Burkardt:2000za,Burkardt:2002hr,Belitsky:2003nz} and contain information about the hadron state, such as the mass and angular momentum \cite{Ji:1996ek,Ji:1994av,Polyakov:2002yz}, have played a pivotal role in the non-perturbative regime of quantum chromodynamics (QCD) both theoretically and experimentally. As a generalization of forward parton distributions (PDFs), which only capture the one-dimensional longitudinal structure, GPDs, proposed as off-forward parton distributions \cite{Ji:1998pc}, leverage nonzero momentum transfer to a hadron. This allows for the inclusion of the skewness parameter $\xi$ describing the fraction of the longitudinal momentum transfer, and the total momentum transfer squared $t,$ which enables probing the 3D structure of the nucleon. To be specific, GPDs are encapsulated as a function of \((x,\xi,t)\), denoted as \(F(x,\xi,t)\), where \(x\) represents the regular parton momentum fraction. In contrast, PDFs, denoted as \(f(x)\), have only one parameter \(x\). The properties (including analyticity, symmetries, sum rules, polynomiality conditions, etc.) and dynamics of GPDs have been well studied over the past three decades and are comprehensively reviewed in \cite{Ji:1998pc,Diehl:2003ny,Belitsky:2005qn}. 

On the other hand, significant efforts from experiments have yielded substantial datasets of exclusive measurements from HERA \cite{aaronMeasurementDeeplyVirtual2008,aktasMeasurementDeeplyVirtual2005,adloffMeasurementDeeplyVirtual2001,H1:2009wnw,chekanovMeasurementDependencesDeeply2009,airapetianBeamhelicityAsymmetryArising2012,airapetianBeamhelicityBeamchargeAsymmetries2012,H1:2005dtp}, Jefferson Lab (JLab) \cite{hattawyExploringStructureBound2019,burkertBeamChargeAsymmetries2021,munozcamachoScalingTestsCrosssection2006,mazouzDeeplyVirtualCompton2007,fonvieilleVirtualComptonScattering2012,georgesDeeplyVirtualCompton2022}, and are expected from the upcoming Electron-Ion Collider (EIC) \cite{Accardi:2012qut,AbdulKhalek:2021gbh}. GPDs can be accessed through exclusive production processes, such as deeply virtual Compton scattering (DVCS) \cite{Ji:1996nm} and deeply virtual meson production (DVMP) \cite{Radyushkin:1996ru}, according to the factorization theorem \cite{Ji:1998xh,Collins:1998be,Collins:1996fb}. However, extracting GPDs, which are high-dimensional quantities, from the amplitudes of these processes remains challenging, as the amplitudes typically only show \(t\) dependence or \(\xi\) and \(t\) dependence from the form factors. This is known as the inverse problem. In addition, the advent of the large momentum effective theory (LaMET) \cite{jiPartonPhysicsEuclidean2013,Ji:2014gla,jiLargemomentumEffectiveTheory2021} expanded the scope of lattice calculations and shed light on the first principle calculation of GPDs \cite{Alexandrou:2021jok,constantinouPartonDistributionsLatticeQCD2021,linNucleonHelicityGeneralized2022,linNucleonTomographyGeneralized2021}. 

Considering all these constraints, the program to parameterize GPDs through universal moment parameterization (GUMP) \cite{Guo:2022upw,Guo:2023ahv} is proposed. This program aims to perform a global analysis by combining lattice calculations, PDFs, and experimental measurements for the form factors, in order to solve the inverse problem. In GUMP, an infinite number of conformal moments of GPDs are parameterized with a finite number of free parameters (called universal parametrization). These parameters are expected to fit the number of different constraints on the GPDs, allowing the GPDs to be fully determined and reconstructed using Mellin-Barnes integrals \cite{muellerComplexConformalSpin2006,Muller:2014wxa}. 
However, the mathematical correlations between the formal summation and the Mellin-Barnes integral remain subtle and unclear. We also observed that the reconstructed GPDs do not vanish at \(|x| = 1\) in previous literature \cite{Guo:2022upw}. Furthermore, developing a framework to convert the multiple summations into multiple Mellin-Barnes integrals is crucial for accurately incorporating the next-to-leading order (NLO) evolution in experimental observables. Thus, a solid mathematical treatment is needed to resolve these issues and support future programs.

The paper is structured as follows. In Section \ref{section2}, we review the Mellin moment expansion and conformal moment expansion of PDFs and GPDs in a logical approach. We heuristically argue the relationship between the formal summation and the Mellin-Barnes integral, connecting them with the extensions of \(\delta^{(n)}(x)\) and conformal partial wave functions \(p_n(x,\xi)\). In Section \ref{section3}, we study the asymptotic expansion of the Mellin-Barnes integral in the complex conformal moment plane and find an approximate formula for $F(x,\xi,t)$ when $x>\xi$. More importantly, we obtain an asymptotic condition for the conformal moments of GPDs to ensure $F(x,\xi,t)$ vanish at \(|x| = 1\). In Section \ref{section4}, we address the problem of double summations when considering the scale evolution operator $E_{jk}^{ii'}(\xi, Q, \mu_0)$ containing non-trivial mixing in the conformal space at NLO \cite{kumerickiFittingProcedureDeeply2008, Muller:2013jur}. This involves the presence of non-diagonal terms in $E_{jk}^{ii'}(\xi, Q, \mu_0)$ between $j$ and $k$ indices. We transform the double summations into double Mellin-Barnes integrals using a novel method to analytically continue the summation with variable bounds, followed by the application of the Sommerfeld-Watson transformation. The method can easily be generalized to handle multiple summations.

\section{Reconstructing PDFs and GPDs from their moments}
\label{section2}
In this section, we provide some new insight on existing Mellin moment and conformal moment expansions for PDFs and GPDs, respectively \cite{Guo:2022upw,Guo:2023ahv,muellerComplexConformalSpin2006,Muller:2014wxa}. We reconstruct the inverse Mellin transformation and Mellin-Barnes integral from the formal summations by the extensions of $\delta^{(n)}(x)$ and the conformal partial wave functions $p_n(x,\xi)$. That is, we begin with the case of PDFs, which possess a well-established inverse Mellin transformation. Through the anatomy of the inverse Mellin transformation, we can convert it into a summation using the Sommerfeld-Watson transformation, as briefly reviewed in Appendix \ref{Sommerfeld}. During this process, we identify an extension of \(\delta^{(n)}(x)\), denoted as \(\mathfrak{D}^n(x)\). This extension facilitates the formulation of a formal summation, involving a ``formal" manipulation. For the case of GPDs, we employ the reverse procedure. Initially, we articulate the formal summation. Subsequently, by determining the appropriate extension for the conformal partial wave function \(p_n(x, \xi)\) as the Schl\"afli integral, we arrive at the Mellin-Barnes integral.

\subsection{Reconstruction of PDFs through the Dirac delta functions}

We begin with the Mellin moment expansion of PDFs, a technique emerging from the analysis of deep inelastic scattering and extensively utilized in quantum field theory. This method finds applications ranging from factorization via operator product expansion (OPE) \cite{White:2001pu,Collins:1981uw,Wilson:1972ee} to the Dokshitzer-Gribov-Lipatov-Altarelli-Parisi (DGLAP) equations \cite{Altarelli:1977zs,Dokshitzer:1977sg,Gribov:1972ri}. In Mellin moment space, the DGLAP equations for PDFs become simple multiplicative expressions that can be easily solved to all orders in perturbation theory. The Mellin moment of PDFs $f(x)$ is defined as
\begin{equation}
  f_n:=\int_{0}^1 \mathrm{~d} x x^{n-1} f(x)\ ,
  \label{mellinmoment}
\end{equation}
where $n=1, 2, ...$ are positive integers. One can reconstruct the PDFs via the inverse Mellin transformation
\begin{equation}
  f(x)=\frac{1}{2 \pi i} \int_{c-i \infty}^{c+i \infty} x^{-j} f_j \mathrm{~d} j\ ,
  \label{invMellinTran}
\end{equation}
where $f_j$ is the analytic continuation of $f_n$ in the complex $j$-plane. 
Supposing that \( f_j \) is a meromorphic function on the complex plane with right bounded poles, we can always choose a sufficiently large \(c\) such that the vertical line \((c-i\infty, c+i\infty)\) is to the right of all the poles ($f_\infty=0)$. 

\begin{figure}[h]
  \centering     
  \includegraphics[width=60mm]{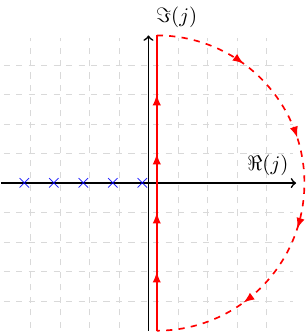}\quad\quad\quad\quad\quad\quad\quad\quad
  \includegraphics[width=60mm]{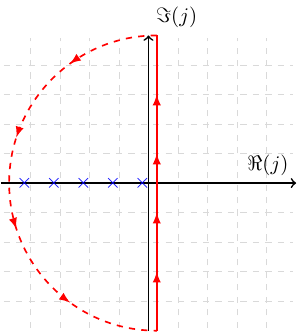}
  \caption{Hypothetical integration contour for $|x|\ge 1$ (left) and $|x|< 1$ (right).}
  \label{contourfigxg1}
\end{figure}

The physical condition \( f(x) = 0 \) only for \( x \ge 1 \) is satisfied as long as asymptotically, 
\begin{equation}
  f_j =\smallO\left( \frac{1}{j}\right) \quad \text{as }  j \to \infty\ .
  \label{conditiononmmcomplex}
\end{equation}
In this case, we can construct a hypothetical semicircle with radius \( R \to \infty \) on the right half-plane, connecting with the vertical line \((c-i\infty, c+i\infty)\), contouring no poles, and there is no contribution to the contour integral on the semicircle. By Cauchy's theorem, the original integral is $0$. The integration contours are depicted in Fig. \ref{contourfigxg1}. Note that the small-$\smallO$ symbol we employ in this paper is slightly different from its strict mathematical definition. It means the function can only decay polynomially faster than \( {1}/{j} \); it cannot decay exponentially, as this would cause the support smaller than $[0,1].$ 
As we can see in (\ref{invMellinTran}), there will be a rescaling of \(x\) if there is an exponentially decaying term in \(f_j\).
In principle, there are many choices of functions to parametrize the Mellin moment \(f_j\). In table \ref{tableofpdf}, we show some models of Mellin moment \(f_j\) and the corresponding PDFs \(f(x)\). 
\begin{table}[h]
\begin{TAB}(r,1cm,1cm)[5pt]{|c|c|}{|c|c|c|c|c|}
  Mellin moment $f_j$  & PDF $f(x)$                      \\ 
  $(j-j_0)^{-\alpha},\, \alpha>1$ & $(-1)^{\alpha-1} {x^{-j_0} \ln ^{\alpha-1}(x)\theta(1-|x|)}/{\Gamma(\alpha)} $    \\ 
  $B(a+j, b+1),\, b>0$ & $f(x)=x^a(1-x)^b \theta(1-|x|)$ \\ 
  $c^{-j} j^{-\alpha},\, \alpha>1$       & $(-1)^{\alpha-1} {\ln ^{\alpha-1}(c x)\theta({1}/{c}-|x|)}/{\Gamma(\alpha)} $ \\
  $c^{-j} B(a+j, b+1),\, b>0$  & $(c x)^a(1-c x)^b \theta({1}/{c}-|x|)$
\end{TAB}
\caption{Some basic models of Mellin moments \(f_j\) and the corresponding PDFs \(f(x)\). The first row represents the simplest polynomial decaying moment, the second row represents the Beta function moment, which is generally used in parameterization \cite{graudenzMellinTransformTechnique1996,stratmannGlobalAnalysisPolarized2001} for its simple PDFs $f(x)=x^a(1-x)^b\theta(1-|x|)$. Note that \(\theta(1 - |x|)\) here ensures that the PDFs vanish when \(|x| \ge 1\). The third and fourth rows represent variations of the first and second rows. As observed, the presence of an exponential decay term results in the PDFs having a more restricted support $[0,1/c].$}
\label{tableofpdf}
\end{table}

Observing  (\ref{invMellinTran}), one can rewrite it in the following form
\begin{align}
  f(x) & =\frac{1}{2 \pi i} \int_{c-i \infty}^{c+i \infty} f_j x^{-j} d j\ ,\\
  \label{manipulation}& =-\frac{1}{2 i} \oint_{C_j} \frac{f_j}{\sin \pi j}\frac{\Gamma(j)\sin{\pi j}}{\pi}\frac{x^{-j}}{\Gamma(j)} d j\ ,
\end{align}
where \(c\) is chosen to lie between 0 and 1, under the assumption that \(f_j\) has no poles with a real part exceeding \(c\), and the contour \(C_j\) encloses all the poles at \(j \in \mathbb{N}^+\), which are artificially introduced by the \(1/\sin{(\pi j)}\) term. Furthermore, we define
\begin{equation}
  \mathfrak{D}^j(x) = \frac{\Gamma(j+1) \sin (\pi[j+1])}{\pi} x^{-j-1},\quad x>0\ ,
  \label{invMellinkernel}
\end{equation}
which is essentially the integral kernel of the inverse Mellin transform, up to a factor. Inserting (\ref{invMellinkernel}) back into (\ref{manipulation}), we can further get
\begin{align}
    f(x) &=-\frac{1}{2 i} \oint_{C_j} \frac{1}{\sin \pi j}\frac{f_j \mathfrak{D}^{j-1}(x) }{\Gamma(j)} d j\ ,\\
    &=\sum_{n=1}^{\infty}(-1)^{n-1} \frac{1}{(n-1)!} f_n \mathfrak{D}^{n-1}(x)\ ,\\
    \label{formalsumorigin}&=\sum_{n=0}^{\infty}\frac{(-1)^{n}}{n!} f_{n+1} \mathfrak{D}^{n}(x)\ .
\end{align}
We applied the Sommerfeld-Watson transformation inversely in the second equality. 

However, the validity of the transformation here is subtle, rendering the expansion merely formal as we will see later. If we consider $x^m$ as a {\it basis} functions, it appears that $f(x)$ is expanded in the {\it dual basis} functions $\mathfrak{D}^n(x).$ We can verify the orthogonal relation between $\mathfrak{D}^n(x)$ and $x^m$ for integers \(n\) and $m$,
\begin{align}
    \int dx~ \mathfrak{D}^{n}(x) x^m 
    &= \frac{n! \sin (\pi[n+1])}{\pi} \int dx~ x^{-n+m-1}\ , \\
    &= \frac{n! \sin (\pi[n+1])}{\pi(m-n)}x^{m-n}\ ,\\
    &= 
    \begin{cases}
      0 & \text{if } m \ne n\ , \\
      (-1)^n n! & \text{if } m = n\ .
    \end{cases}
\end{align}
For the case \(m = n\), L'Hôpital's rule has been applied. This result indicates that \(\mathfrak{D}^{n}(x)\) behaves as the derivatives of the Dirac delta function \(\delta^{(n)}(x)\), which satisfies:
\begin{equation}
    \int_{-1}^1 dx~ \delta^{(n)}(x) x^m = (-1)^n n! \delta_{mn}\ .
    \label{normcondi}
\end{equation}
Thus, we can identify \(\mathfrak{D}^{n}(x)\) as \(\delta^{(n)}(x)\) due to their congruent properties. Therefore, equation (\ref{formalsumorigin}) can be interpreted as the formal summation shown in \cite{Guo:2022upw}:
\begin{equation}
    f(x) = \sum_{n=0}^{\infty} \frac{(-1)^n}{n!} f_{n+1} \delta^{(n)}(x)\ .
    \label{formalsummationofpdf}
\end{equation}
Note that the order of summation and integration cannot actually be interchanged when acting on (\ref{formalsumorigin}) or (\ref{formalsummationofpdf}) with \(x^n\), making such manipulation merely ``formal." Nonetheless, both approaches yield the \((n+1)\)-th Mellin moment \(f_{n+1}\). In this context, \(x^m\) and \(\delta^{(n)}(x)\) serve as the basis and dual basis in their respective function spaces, satisfying the orthogonal condition (\ref{normcondi}). The support of \(f(x)\) should be \([0, 1]\), whereas the support of \(\delta^{(n)}(x)\) or \(\mathfrak{D}^n(x)\) is \(\{0\}\). This discrepancy contributes to the divergence observed in (\ref{formalsummationofpdf}). Specifically, for the formal summation (\ref{formalsummationofpdf}) with singular support at \(\{0\}\), no non-trivial Mellin moment \(f_n\) can converge: convergence only happens when the PDF has support only at $x=0$, namely a delta function. In fact, we can use functional analysis and introduce the concepts of distributions and analytic functionals \cite{tsunoAnalyticFunctionalsDistributions1970} to rigorously formalize the summation (\ref{formalsummationofpdf}). We omit the detailed mathematical exposition of these concepts here to maintain our focus on the physical aspects.

Note that we can also express equation (\ref{invMellinkernel}) as follows:
\begin{align}
    \mathfrak{D}^j(x) &= \frac{\Gamma(j+1)}{2\pi i}\left(e^{i\pi j} - e^{-i\pi j}\right) x^{-j-1}\ , \\
    \label{deltaexten}&= -\frac{\Gamma(5/2 + j)\Gamma(1 + j)}{\Gamma(1/2) \Gamma(2 + j)} \frac{1}{2\pi i} \oint_{-1}^1 du \frac{\left(u^2 - 1\right)^{j + 1}}{x^{j + 1}}\ ,
\end{align}
where $u$ is a complex variable and the contour encircles the interval \(u \in [-1, 1]\). 
The above integral expression is useful for analyzing GPD expansion in the next subsection.

\subsection{Reconstructing GPDs from conformal moments and partial wave functions}

For GPDs \(F(x, \xi, t)\), the reconstruction from 
moments can in principle be made in parallel to the previous subsection. One can start from Mellin moments $F_j(\xi, t)$, and obtain $F(x, \xi, t)$ through a similar inverse Mellin transformation. However, this straightforward generalization obscures a very important analytical property 
of GPDs. { The variables of GPDs are defined as
\begin{equation}
x=\frac{2k^{+}}{P^{+}+P'^{+}}\ , \quad \xi=-\frac{\Delta^{+}}{P^{+}+P'^{+}}\ , \quad t=\Delta^2\ , \quad\Delta\equiv P'-P\ ,
\end{equation}
where $P$ and $P'$ are momenta of the incoming and outgoing nucleon, $k$ is the momentum carried by the parton.}

\begin{figure}[h]
  \centering     
  \includegraphics[width=150mm]{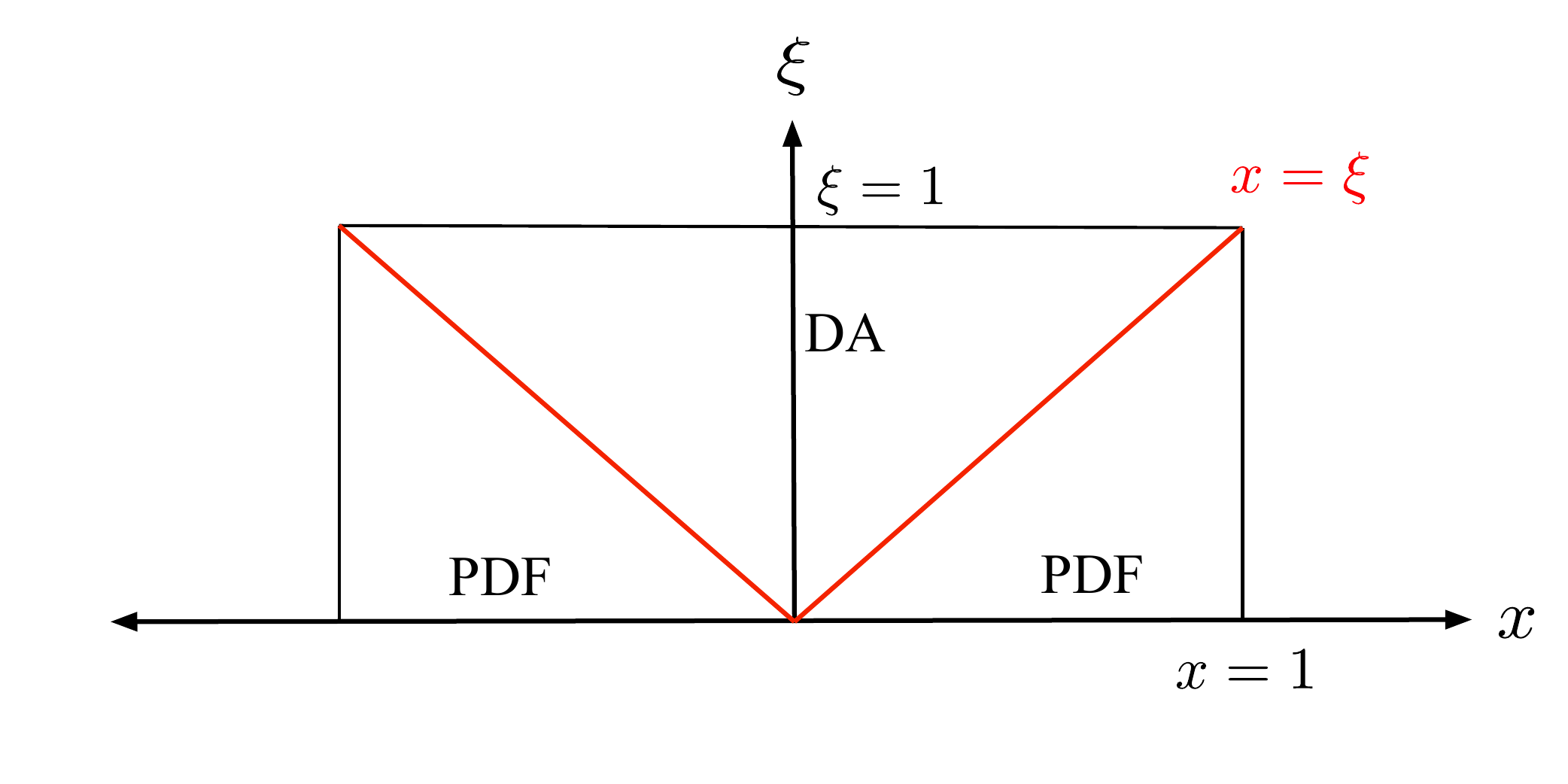}
  \vspace{-1cm}
  \caption{``Phase diagram" of GPDs, showing different kinematical regions on the $(x,\xi)$ plane. This figure is sourced from \cite{Guo:2022upw}.}
  \label{phasediagram}
\end{figure}

\begin{figure}[h]
  \centering     
  \includegraphics[width=0.4\textwidth]{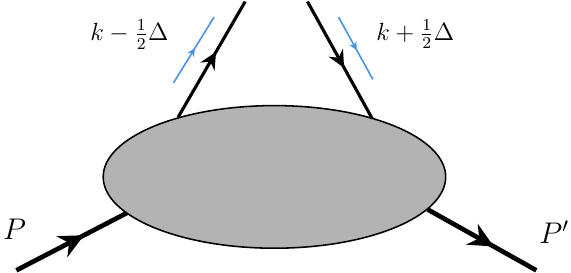}\quad
  \includegraphics[width=0.4\textwidth]{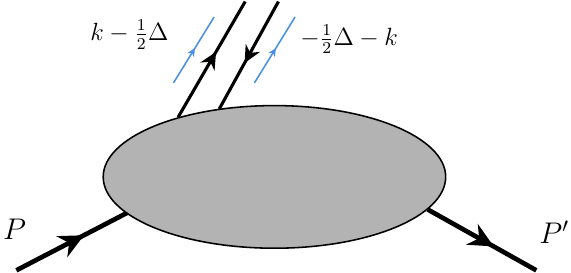}
  \caption{{Kinematics for GPDs in PDF-like (left) and DA-like (right) regions.}}
  \label{kinematicsgpd}
\end{figure}

Indeed, the physics of GPD functions can be divided into two separated regions as illustrated in the ``phase diagram" Fig. \ref{phasediagram}. These are the PDF-like region (\(x \ge \xi\) and \(x \le -\xi\)) and the distribution amplitude (DA)-like region (\(-\xi < x < \xi\)), also referred to as the DGLAP and Efremov-Radyushkin-Brodsky-Lepage (ERBL) regions in the literature, respectively. These two regions connect at the lines \(x = \pm \xi\), where the GPDs are continuous but not analytic. This property of GPDs is ensured by the factorization theorem \cite{Collins:1998be}.  Physically, the non-analyticity arises from the differing interpretations in these two regions as shown in Fig. \ref{kinematicsgpd}. In the PDF-like region, GPDs represent the probability amplitude for the emission and absorption of a quark with different momentum fractions, analogous to the interpretation of PDFs. In contrast, in the DA-like region, GPDs represent the probability amplitude of a meson-like entity within the hadron, similar to the DA in meson production. At the lines \(x = \pm \xi\), one of the two partons has zero momentum, resulting in a ``phase transition." 

One can in principle reconstruct the above phase diagram through polynomiality condition \cite{Ji:1998pc} on the GPD moments. However, there is a simpler way to do this:
Instead of expanding formally in terms of Mellin moments and $\delta^{(n)}(x)$ with singular support of $\{0\}$, one can introduce new class of
moments through orthogonal polynomials in $x\in [-1,1]$ and expand the GPD in terms of the dual basis of these polynomials 
defined in the larger support region $x\in[-\xi, \xi]$, which reduces to the $\delta^{(n)}(x)$ in the forward limit. The dual basis
has the same analytic property as GPDs that they are continuous but not analytic at $x=\pm \xi$. In this way, the phase-diagram of GPD is naturally incorporated from the beginning.  

Although any choices of orthorgonal polynomials shall work, we follow \cite{muellerComplexConformalSpin2006} to use the conformal moments, as a generalization of Mellin moments, to perform the conformal partial wave expansion  or conformal OPE \cite{Muller:2013jur}. The advantage of using conformal moments is that, when considering the DGLAP equations for GPDs, the convolution between the leading-order (LO) evolution kernel and GPDs can be diagonalized in the conformal moment space \cite{Belitsky:1997pc,Guo:2022upw}. However, at NLO and even higher orders, non-diagonal terms persist, causing non-trivial mixing of conformal indices between the evolution kernel and conformal moments, as we will examine in Sec. \ref{section4}. In that case, double summations and double complex-plane integrals will be involved, which makes the above choice
less obvious.

The conformal moment of GPD $F(x, \xi, t)$ is defined as 
\begin{equation}
  \mathcal{F}_n(\xi, t) = \int_{-1}^1 dx \, c_n(x, \xi) F(x, \xi, t)\ ,
  \label{defconformalmoment}
\end{equation}
where $c_n(x, \xi)$ are Gegenbauer polynomials $C_n^{\frac{3}{2}}({x}/{\xi})$ up to a conventional normalization factor, that is \cite{muellerComplexConformalSpin2006}
\begin{equation}
    c_n(x, \xi)=\xi^n \frac{\Gamma\left(\frac{3}{2}\right) \Gamma(1+n)}{2^n \Gamma\left(\frac{3}{2}+n\right)} C_n^{\frac{3}{2}}\left(\frac{x}{\xi}\right)\ ,
\end{equation}
where $n$ are also positive integers. Especially, it reduces to $x^n$ for Mellin moment in the forward limit, that is $\lim _{\xi \rightarrow 0} c_n(x, \xi)=x^n.$ If we select \(c_n(x, \xi)\) as the {\it basis} of the function space, the corresponding {\it dual basis} \(p_n(x, \xi)\), known as the conformal partial wave function, can be determined through the orthogonality condition:
\begin{equation}
    \int_{-\xi}^{\xi} p_m(x, \xi) c_n(x, \xi) \, dx = (-1)^n \delta_{mn}\ .
    \label{orthoconditiongpd}
\end{equation}
Note that the integral bounds are \(x \in [-\xi, \xi]\). This constraint is necessary because this relationship holds only for \(x/\xi \in [-1, 1]\), owing to the presence of the weighting function \(w(x/\xi) = [1 - (x/\xi)^2]\) in the orthogonal relation for the Gegenbauer polynomials:
\begin{equation}
    \int_{-\xi}^\xi C_m^{\frac{3}{2}}\left(\frac{x}{\xi}\right)C_n^{\frac{3}{2}}\left(\frac{x}{\xi}\right) w\left(\frac{x}{\xi}\right)  d x=\xi\frac{\Gamma (n+3)}{\left(n+\frac{3}{2}\right) \Gamma (n+1)}\delta_{mn}\ .
    \label{orthorelationge}
\end{equation}
The conformal partial wave function \(p_n(x, \xi)\) is then defined as follows: 
\begin{equation}
  p_n(x, \xi) = (-1)^n \xi^{-n-1} \frac{2^n \Gamma\left(\frac{5}{2}+n\right)}{\Gamma\left(\frac{3}{2}\right) \Gamma(3+n)} \left[1 - \left(\frac{x}{\xi}\right)^2\right] C_n^{\frac{3}{2}}\left(\frac{x}{\xi}\right) \theta(\xi - |x|) \quad \text{for } |x| < \xi\ ,
  \label{partialwaveori}
\end{equation}
where \(\theta(\xi - |x|)\) is the Heaviside step function, ensuring the function's definition is restricted to the interval \(|x| < \xi\). 

In analogy with Eqs. (\ref{formalsumorigin}) and (\ref{formalsummationofpdf}), the formal expansion/summation for GPDs now reads:
\begin{align}
  F(x,\xi) = \sum_{n=0}^{\infty} (-1)^n \mathcal{F}_n(\xi, t) p_{n}(x,\xi)\ .
  \label{formalsummationofgpd}
\end{align}
The above properties of $p_n(x, \xi)$ result in the exclusion of information for \(x > \xi\) and \(x < -\xi\), leading to the divergence of the formal summation. The reduction of support necessitates a stronger decaying condition on the conformal moments $\mathcal{F}_n$ of GPDs to make the formal summation converge, as discussed in the case of PDFs. More specifically, if GPDs are non-vanishig only in the DA-like region, the resulting moments will make the above expansion convergent. Any non-vanishing contributions in the PDF-like region will generate conformal moment series rendering the above expansion divergent. When convoluting the formal summation (\ref{formalsummationofgpd}) with \(c_n(x, \xi)\), we anticipate it to yield the \(n\)-th conformal moment \(\mathcal{F}_j(\xi, t)\). Nonetheless, the legitimacy of interchanging between the integral and summation in this context is subtle, rendering this also a ``formal'' manipulation.
An appropriate resummation is needed to extract physical results. There are several resummation techniques proposed in the literatures \cite{Belitsky:1997pc,Mankiewicz:1997uy,Manashov:2005xp,Shuvaev:1999fm,Polyakov:2002wz}. Here, we reconstruct GPDs using the Mellin-Barnes integral suggested in \cite{muellerComplexConformalSpin2006, Muller:2014wxa}, which is based on the Sommerfeld-Watson transformation. In this case, one has to analytically continue the moments and $p_n(x,\xi)$ to complex $j$ plane. 

It is easy to see that in the forward limit, the above formalism reduces to that for PDFs. 
Indeed, the conformal partial wave function $p_n(x, \xi)$ can also be written as
\begin{equation}
  p_{n}(x,\xi)=\frac{\Gamma(\frac{5}{2}+n)}{n!\Gamma(\frac{1}{ 2}) \Gamma(2+n)} \int_{-1}^1 d u\left(1-u^2\right)^{n+1} \delta^{(n)}(x-u \xi)\ ,
  \label{partialwave}
\end{equation}
which makes the orthogonality condition (\ref{orthoconditiongpd}) easy to verify using:
\begin{align}
    \label{deltage}\int_{-\xi}^{\xi} \delta^{(n)}(x - u \xi) C_n^{\frac{3}{2}}\left(\frac{x}{\xi}\right)  dx = (-1)^n \left. \frac{d^n}{dx^n} C_n^{\frac{3}{2}}\left(\frac{x}{\xi}\right) \right|_{x = u \xi} =(-1)^n \xi^{-n} 2^n \frac{\Gamma\left(\frac{3}{2} + n\right)}{\Gamma\left(\frac{3}{2}\right)}\ .
\end{align}
Note that due to the presence of \(\delta^{(n)}(x - u \xi)\), the support of \(p_n(x, \xi)\) is also confined to \([- \xi, \xi]\). In the forward limit, \(\lim _{\xi \rightarrow 0} p_{n}(x,\xi)= \delta^{(n)}(x)/n!\), which reduces to the case for PDFs.

However, to make the analytical continuation of $p_n(x,\xi)$, we need another 
expression, analogous but not identical to the extension of \(\delta^{(n)}(x)\) to \(\mathfrak{D}^j(x)\).  It turns out that we can generalize (\ref{deltaexten}) by employing the Schl\"afli integral \cite{Schlafli} to properly address poles and branch cuts, as detailed in references \cite{muellerComplexConformalSpin2006, Muller:2014wxa}. Specifically, in our context, this involves modifying \(x\) to \(x + u\xi\) in (\ref{deltaexten}),
\begin{equation}
  p_j(x, \xi)=-\frac{\Gamma(5 / 2+j)}{\Gamma(1 / 2) \Gamma(2+j)} \frac{1}{2  \pi i} \oint_{-1}^1 d u \frac{\left(u^2-1\right)^{j+1}}{(x+u \xi)^{j+1}}\ .
  \label{Schlafliint}
\end{equation}
The integral can be expressed in terms of hypergeometric functions after carefully considering the poles and branch cuts. We summarize the results as follows:
\begin{align}
	\label{gpdspartialwave}&p_j(x>\xi, \xi)=\frac{\sin (\pi[j+1])}{\pi} x^{-j-1}{ }_2 F_1\left(\left.\begin{array}{c}
		(j+1) / 2,(j+2) / 2 \\
		5 / 2+j
		\end{array} \right\rvert\, \frac{\xi^2}{x^2}\right)\ ,\\
	\label{gpdsda}&p_j(|x| \leq \xi, \xi)=\frac{2^{j+1} \Gamma(5 / 2+j) \xi^{-j-1}}{\Gamma(1 / 2) \Gamma(1+j)}(1+x / \xi)_2 F_1\left(-1-j, j+2,2 \left\lvert\, \frac{\xi+x}{2 \xi}\right.\right)\ ,\\
	&p_j(x \leq-\xi, \xi)=0  \ .
\end{align}
It has been verified that \(p_j\) is continuous at \(x = \pm \xi\), but not analytic at these points.
In the forward limit, equation (\ref{gpdspartialwave}) reduces to equation (\ref{invMellinkernel}). In the DA-like region, equation (\ref{gpdsda}) reduces to equations (\ref{partialwave}) and (\ref{partialwaveori}). Observe that for integer \(j\) and \(x = \pm \xi\), the conformal partial wave function \(p_j(x,\xi)\) is  zero.  

To do the resummation, applying the Sommerfeld-Watson transformation to equation (\ref{formalsummationofgpd}), the Mellin-Barnes integral in Ref. \cite{muellerComplexConformalSpin2006, Muller:2014wxa} is obtained:
\begin{equation}
  F(x, \xi, t) = \frac{1}{2i} \int_{c-i\infty}^{c+i\infty} dj \frac{p_j(x, \xi)}{\sin(\pi[j+1])} \mathcal{F}_j(\xi, t)\ .
  \label{Mellinbarnes}
\end{equation}
In the forward limit, it reduces to the inverse Mellin transformation as the PDFs. Furthermore, when \(x =\pm \xi\), the GPDs are {\it not} zero, as the evaluation of the integral along the imaginary axis ensures that the sine term in (\ref{gpdspartialwave}) remains non-zero. Further studies on its analytic properties will be conducted in the next section. Note that in this resummation, GPDs $F(x,\xi,t)$ are zero for \(x < -\xi\). This situation is analogous to quark PDFs, where we set \(f(x) = 0\) for \(x < 0\), with the negative \(x\) region included in the antiquark PDFs. Since GPDs do not inherently distinguish between quark and antiquark components, they are formulated explicitly as a linear combination of various contributions: quark GPDs, antiquark GPDs, and an additional DA term, as discussed in \cite{Guo:2023ahv}.

We would like to emphasize again that the strategy in this subsection is quite general. In principle, any complete set of polynomials can serve as basis functions with the argument \(x/\xi\) to expand the GPDs and derive their respective moments. By applying the orthogonality condition, we can find the corresponding dual basis functions with a restricted support, which is smaller than $[-1,1]$. Due to discrepancies of the supports, the formal summation will be divergent. However, after identifying a suitable generalization of the dual basis functions, we can employ the Sommerfeld-Watson transformation to resum the moments and reconstruct the GPDs accurately over the entire range. 

\section{Mellin-Barnes integrals and asymptotic behavior of GPDs}
\label{section3}
In this section, we study the analytic properties of the Mellin-Barnes integral on the complex \(j\)-plane. We begin with the concept of residue at infinity, then use this concept to examine the inverse Mellin transformation (\ref{invMellinTran}) and the Mellin-Barnes integral (\ref{Mellinbarnes}) by studying their asymptotic behaviors.  Next, we derive the asymptotic conditions on GPDs in the complex plane and test the formulation with a minimal example of GUMP. The asymptotic conditions are of great importance, as it ensures that the GPDs vanish at \(|x| = 1\), a property that was violated in previous parameterizations \cite{Guo:2022upw}. Finally, we derive an approximate formula for GPDs when \(x > \xi\).
\subsection{Residue at infinity}
It is useful to introduce the extended complex plane as \(\mathbb{C}^* = \mathbb{C} \cup \{\infty\}\), which is compact and isomorphic to the Riemann sphere. Define the residue at infinity for a function \(f(z)\) as \cite{ablowitzComplexVariablesIntroduction2003}
\begin{equation}
  \operatorname{Res}(f, \infty) = -\operatorname{Res}\left(\frac{1}{z^2} f\left(\frac{1}{z}\right), 0\right)\ .
\end{equation}
For a meromorphic function on the Riemann sphere, the sum of the residues (including the one at infinity) must be zero. This can be regarded as another definition of the residue at infinity:
\begin{equation}
  \operatorname{Res}(f, \infty) = -\sum_{k=1}^n \operatorname{Res}(f, a_k)\ ,
\end{equation}
where $a_k$ are poles of $f(z)$ in $\mathbb{C}.$ 
We start with the inverse Mellin transformation (\ref{invMellinTran}) reconstructing PDFs to elaborate on this concept. We have studied the properties of this integral when \(|x| \ge 1\) to ensure it yields zero. When $|x|<1,$ we construct a contour to encircle all the poles as in Fig. \ref{contourfigxg1}, that is
\begin{align}
  f(x)&=\frac{1}{2 \pi i} \int_{c-i \infty}^{c+i \infty} x^{-j} f_j \mathrm{~d} j\ ,\\
  &=\sum_{k=1}^n \operatorname{Res}(x^{-j} f_j, a_k)\ ,\\
  &=-\operatorname{Res}(x^{-j} f_j, \infty)\ .
\end{align}
Denoting the inverse Mellin transformation from $j$ space to $x$ space as $f(x)=\left\{\mathcal{M}^{-1} f_j\right\}(x),$ we have
\begin{equation}
  \left\{\mathcal{M}^{-1} f_j\right\}(x)=-\operatorname{Res}(x^{-j} f_j, \infty)\ .
  \label{invMellinandresidue}
\end{equation}
One can easily verify this for a simple PDF model $f(x)=x^\alpha(1-x)^\beta$ if the Mellin moment $f_j=B(j+\alpha,\beta+1).$

The situation for GPDs is alike when $x>\xi,$ which is in the PDF-like region. From (\ref{Mellinbarnes}) and (\ref{gpdspartialwave}), we have
\begin{equation}
  F(x, \xi, t)  =\frac{1}{2 \pi i} \int_{c-i \infty}^{c+i \infty} d j x^{-j-1}{ }_2 F_1\left(\begin{array}{c}
    (j+1) / 2,(j+2) / 2 \\
    5 / 2+j
    \end{array}\right.\left\lvert\, \frac{\xi^2}{x^2}\right) \mathcal{F}_j(\xi, t)\ ,
    \label{mbintxgxi}
\end{equation}
with \(c\) chosen to be to the right of all the poles. Analyzing this integral, we find that:
\begin{enumerate}
\item When \(x = 1\), the integral should be zero. We impose conditions on \(\mathcal{F}_j(\xi, t)\) such that \({}_2F_1(\dots) \mathcal{F}_j(\xi, t)\) asymptotically decays faster than \(1/j\). Then by adding a right half infinite hemicircle, whose contribution to the integral is zero, the original integral equals the sum of residues in the contour. However, after carefully chosen $c,$ \(\mathcal{F}_j(\xi, t)\) and the hypergeometric function both have no poles there. Therefore, the integral is zero as expected. The asymptotic expansion for the hypergeometric function can be found in \cite{parisAsymptoticsGaussHypergeometric2013},
\begin{multline}
  { }_2 F_1\left(\left.\begin{array}{c}
  (j+1) / 2,(j+2) / 2 \\
  5 / 2+j
  \end{array} \right\rvert\, z:=\frac{\xi^2}{x^2}\right) \sim 2 \sqrt{2} C(z)\left[\frac{2}{1+\sqrt{1-z}}\right]^j\quad \text{as }|j| \rightarrow \infty,\, |\arg j|<\pi,\, z<1\ ,
  \label{asymptoticexpansion}
\end{multline}
where $C(z)=\frac{(\sqrt{1-z})^{1 / 2}}{(1+\sqrt{1-z})^{3 / 2}}$ is a constant for given $z$. It shows that to make \({}_2F_1(\dots) \mathcal{F}_j(\xi, t)=\smallO(1/j)\), as \(|j| \rightarrow \infty, |\arg j|< \pi/2\) when $x=1,$ the conformal moment must satisfy
\begin{equation}
  \mathcal{F}_j =\left(\frac{2}{1+\sqrt{1-\xi^2}}\right)^{-j} \smallO\left(\frac{1}{j}\right) \quad \text{as }|j| \rightarrow \infty, |\arg j|< \frac{\pi}{2}\ .
  \label{decayconditioncomplex}
\end{equation}
Furthermore, we can substantiate the assertion made in the previous section: if we aim to restrict the support of GPDs to \([- \xi, \xi]\) as the support of dual basis functions and ensure the convergence of the formal summation (\ref{formalsummationofgpd}), the asymptotic condition of GPDs will be
\begin{equation}
  \mathcal{F}_j' = \left(\frac{2/\xi}{1 + \sqrt{1 - \xi^2}}\right)^{-j} \smallO\left(\frac{1}{j}\right) \quad \text{as } |j| \rightarrow \infty, \, |\arg j| < \frac{\pi}{2}\ ,
  \label{decayconditioncomplex2}
\end{equation}
which actually represents a rescaling of \(x\) in (\ref{mbintxgxi}). This condition is stricter than (\ref{decayconditioncomplex}), explaining why the formal summation of the moments derived from the GPDs expansion over the range \([-1, 1]\) is typically divergent.

\item When \(x > 1\), the factor \(x^{-j-1} \to 0\) as \(|j| \rightarrow \infty, |\arg j| <  \pi/2\), causing the integrand to decay even faster than in the case for \(x = 1\). Hence, the integral is also zero if it is zero at $x=1$, as anticipated.

\item When \(\xi < x < 1\), similarly, the factor \(x^{-j-1} \to 0\) as \(|j| \rightarrow \infty, \pi/2<|\arg j| <\pi \), and the integral on the left hemicircle is zero. The original integral equals the sum of residues in the contour and is equal to negative of the residue at infinity. That is 
	\begin{align}
		F(x, \xi, t)&=\sum_{k=0}^n\text{Res}\left[ x^{-j-1}{ }_2 F_1\left(\begin{array}{c}
		(j+1) / 2,(j+2) / 2 \\
		5 / 2+j
		\end{array}\right.\left\lvert\, \frac{\xi^2}{x^2}\right) \mathcal{F}_j(\xi, t);\, a_k\right]\ ,\\
		\label{residueatinftygpds}&=-\text{Res}\left[ x^{-j-1}{ }_2 F_1\left(\begin{array}{c}
			(j+1) / 2,(j+2) / 2 \\
			5 / 2+j
			\end{array}\right.\left\lvert\, \frac{\xi^2}{x^2}\right) \mathcal{F}_j(\xi, t);\, \infty\right]\ .
		\end{align}
  This suggests that the GPDs in the PDF-like region are exclusively linked to the asymptotic behaviors of both the hypergeometric function ${}_2F_1(\dots)$ and the conformal moment $\mathcal{F}_j(\xi, t)$ in the complex-$j$ plane.
\end{enumerate}

\subsection{A minimal example of GUMP model}
We then show the numerical calculations of a minimal example of GUMP. Because of the polynomiality condition of conformal moments of the GPDs \cite{Ji:1998pc,Guo:2022upw}, we parametrize the conformal moment at small $\xi$ as \cite{kumerickiDeeplyVirtualCompton2010}
\begin{equation}
  \mathcal{F}_j(\xi, t)=\mathcal{F}_{j, 0}(t)+\xi^2 \mathcal{F}_{j, 2}(t)+\mathcal{O}\left(\xi^4\right)\ ,
  \label{polycondi}
\end{equation}
and the coefficient before the skewness parameter $\xi$ is 
\begin{equation}
    \mathcal{F}_{j, k}(t)= N_{ k} B\left(j+1-\alpha_{ k}, 1+\beta_{k}\right) \frac{j+1-k-\alpha_{ k}}{j+1-k-\alpha_{ k}-\alpha_{ k}'t} \beta(t)\ ,
\end{equation}
in which $N_k,\,\alpha_k,\,\beta_k,\,\alpha_k'$ are parameters determined by global fitting, and the $t$ dependence is incorporated by Regge trajectory term \cite{Regge:1959mz,Regge:1960zc} and the extra residual term $\beta(t)$ \cite{Guo:2023ahv}. For simplicity, we only consider the leading term \(\mathcal{F}_{j,0}\), since $\xi=x_B /\left(2-x_B\right)+\mathcal{O}\left(Q^{-2}\right)$ is small in the high energy limit \cite{kumerickiDeeplyVirtualCompton2010}.We neglect the \(t\) dependence, as it can always be recovered by simply multiplying Regge terms, and focus solely on the \(j\) dependence. We have the minimal model
\begin{equation}
  \mathcal{F}_{j,\text{original}}(\xi,0)= B\left(j+1-\alpha, 1+\beta\right)\ .
\end{equation}
Note that the \(j+1\) index in the conformal moment corresponds to the \(j\) index in the Mellin moment conventionally, so \(\mathcal{F}_{j, \text{original}}\) corresponds to generalization of the PDF \(f(x) = x^{-\alpha}(1-x)^\beta\), which is the most basic ansatz in the CTEQ global analysis \cite{houNewCTEQGlobal2021}. Since this moment does not satisfy the condition (\ref{decayconditioncomplex}), we observe that the GPDs reconstructed from it do not vanish at \(x =  1\) \cite{Guo:2022upw}. However, for relatively not large values of \(\xi\lesssim 0.4\) (most available experimental data fall into this range), the deviation is negligible.

\begin{figure}[h] \centering
  \includegraphics[width=0.6\textwidth]{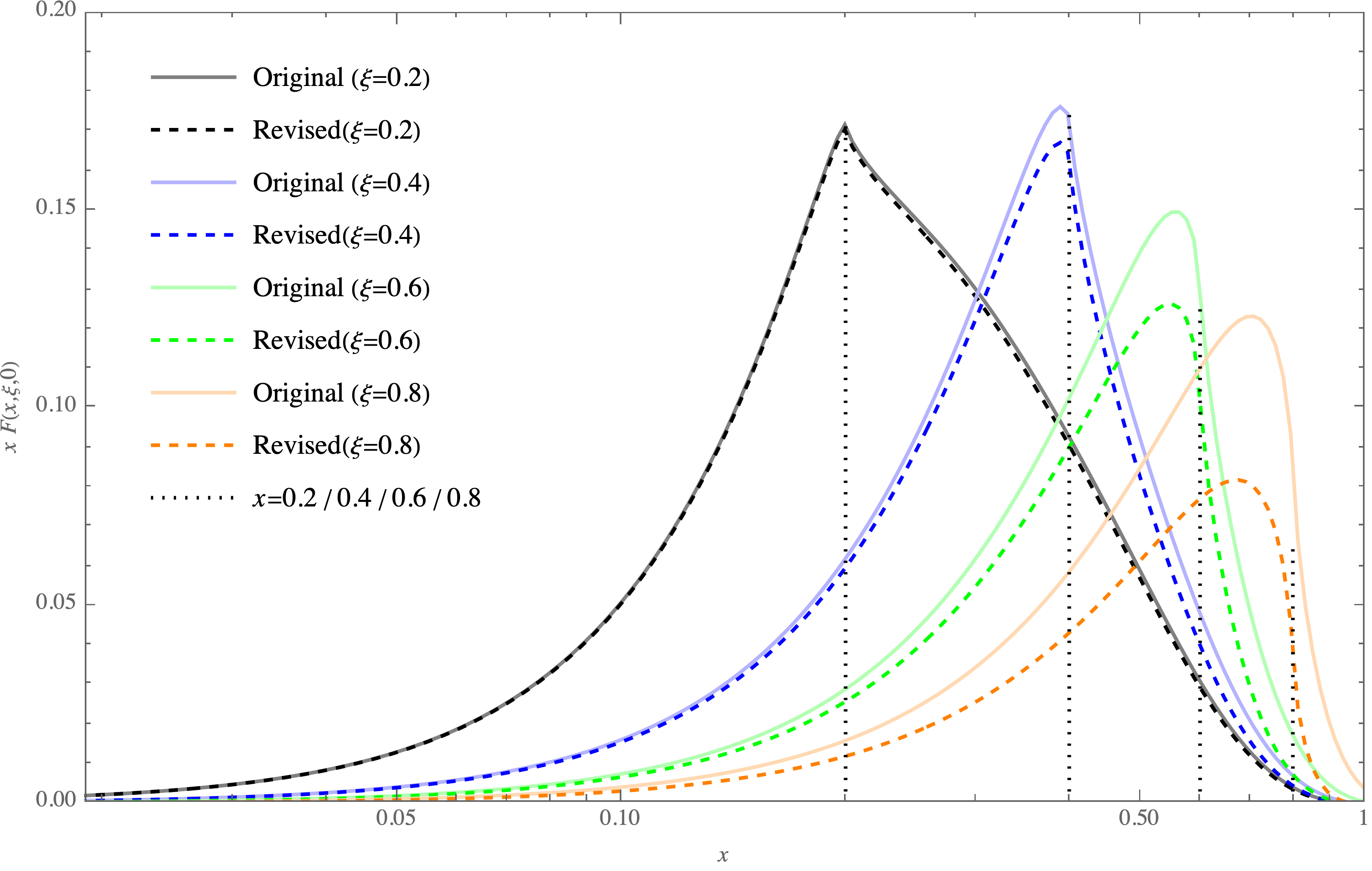}
  \caption{ A plot showing the rescaled GPDs, \(xF(x, \xi, 0)\), versus the parton momentum fraction \(x\) for different skewness parameters \(\xi = 0.2, 0.4, 0.6, 0.8\). The solid lines represent the GPDs reconstructed from original moments, while the dashed lines correspond to the ones reconstructed from revised moments at \(t = 0\). The lines \(x = \xi\) are marked with dotted black lines. The plot demonstrates that for moderate skewness, the deviation between the original and revised moments is negligible, whereas at larger skewness values, the deviation becomes more pronounced, with the GPDs decaying more rapidly as \(x\) approaches 1.}
  \label{originrevise}
\end{figure}

Theoretically, to make it precisely vanish at \(x = 1\), we multiply the moment by the exponential decaying term as specified in (\ref{decayconditioncomplex}). Thus, the moment now becomes
\begin{equation}
  \mathcal{F}_{j,\text{revised}}(\xi,0)= B\left(j+1-\alpha, 1+\beta\right) \left(\frac{2}{1+\sqrt{1-\xi^2}}\right)^{-j-1}\ .
  \label{momentrevised}
\end{equation}
Actually, this revision will not cause much difference for moderate skewness, as shown in Fig. \ref{originrevise}. 
Only when \(\xi\) is large does it bring a noticeable deviation, causing the GPDs to decay faster towards \(x = 1\) and become significantly smaller. { An obvious drawback of the revised moment is that it does not obey the polynomiality conditions. One can remedy this by expanding the above 
$\xi$ factor at $\xi=0$ and truncate to an appropriate order, which can be done analytically by using the Mellin-Barnes technique introduced in Sec. \ref{finiteanalytic} to deal with the finite summation. The explicit formula for the moment that satisfies the polynomiality condition is given by
\begin{multline}
	\mathcal{F}_{j,\text{polynomial}}(\xi,0)=B\left(j+1-\alpha, 1+\beta\right)\left(\frac{2}{1+\sqrt{1-\xi^2}}\right)^{-j-1}-2^{-j-4} (j+1) \xi ^{j+3} \left\{\left[1+(-1)^j\right] \xi  \, _3\tilde{F}_2\left(1,\frac{3}{2},2;\frac{4-j}{2},\frac{j+6}{2};\xi ^2\right)\right.\\
	\hspace{5cm}\left.+\left[1-(-1)^j\right] \, _3\tilde{F}_2\left(1,1,\frac{3}{2};2-\frac{j+1}{2},\frac{j+1}{2}+2;\xi ^2\right)\right\}B\left(j+1-\alpha, 1+\beta\right)\ .
\end{multline}
It can be verified that for integer values of $j$, the formula above produces a polynomial of $\xi$ that satisfies the polynomiality condition. As discussed in the previous subsection, since the GPDs only have to do with the asymptotic behavior of the moment, truncation is not necessary since it will not affect the GPDs in the PDF-like region, and in the DA-like region, the difference remains negligible for the same reason. Therefore, we can consistently use the asymptotic moment (\ref{momentrevised}) and disregard the polynomiality condition, as it will not result in any noticeable  deviation.} 
Besides, we have verified the consistency of the Mellin-Barnes integral formulation. Starting from the GPDs \(F(x, \xi, t)\) reconstructed from (\ref{momentrevised}) and (\ref{Mellinbarnes}), we transform it again using the definition of the conformal moment (\ref{defconformalmoment}). The result yields exactly the same conformal moment we used initially.

\subsection{An approximation formula for $F(x,\xi,t)$ when $x>\xi$}
\label{sectionapprox}
\begin{figure}[h] \centering
  \includegraphics[width=0.6\textwidth]{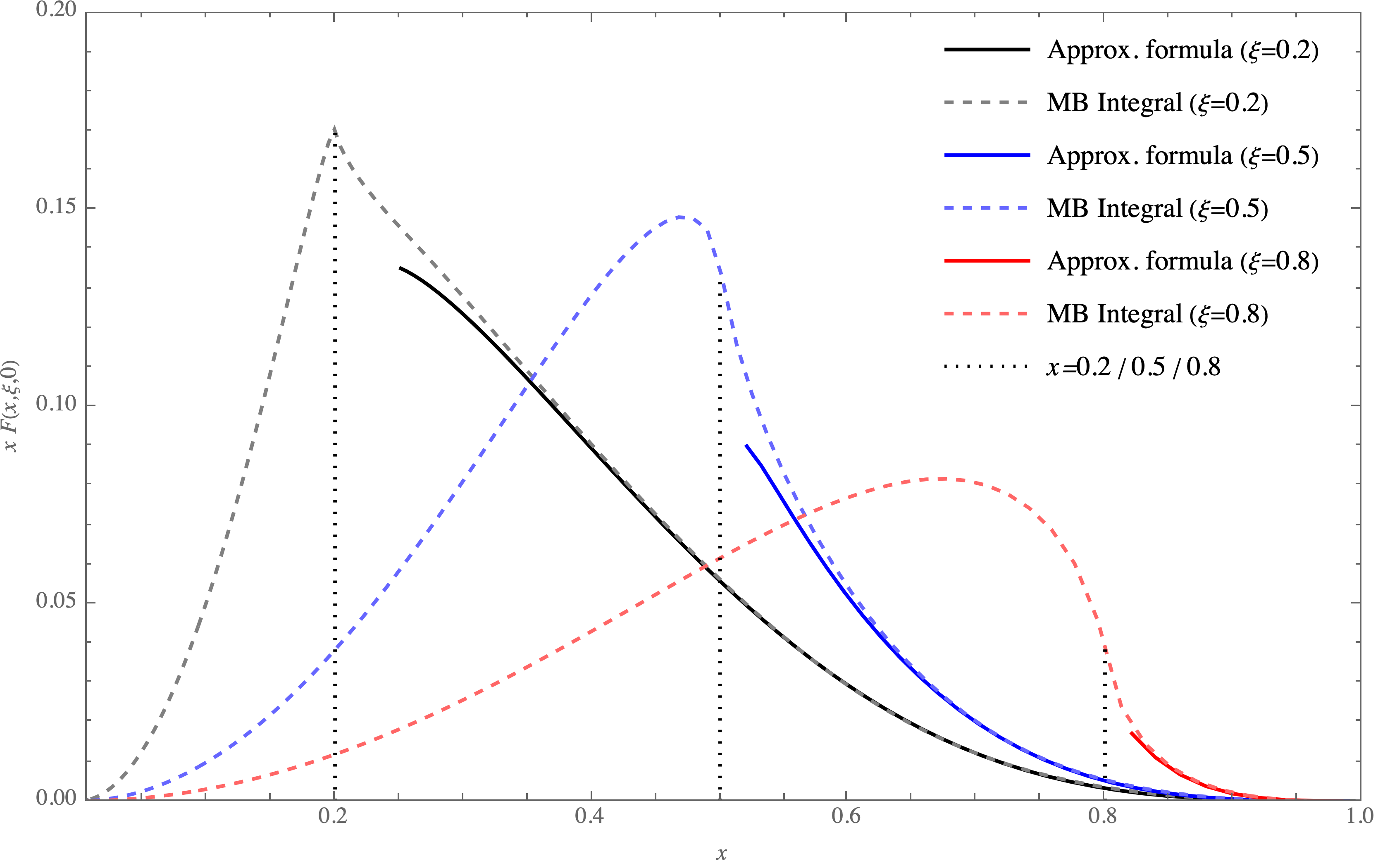}
  \caption{A plot showing the comparison between the approximation formula (solid lines) for PDF-like GPDs (\(x > \xi\)), derived from the asymptotic expansion of hypergeometric functions in conformal partial waves, and the numerical reconstruction of GPDs using the Mellin-Barnes (MB) integral (dashed lines), plotted against the parton momentum fraction \(x\). The lines at \(x = \xi\) are indicated by dotted black lines. The plot shows that the approximation formula is quite accurate for values of \(x\) that are not too close to \(\xi\), with the deviation becoming minimal as \(x\) moves further away from \(\xi\).}
  \label{approxmbint}
\end{figure}

Using the asymptotic expansion of the hypergeometric function (\ref{asymptoticexpansion}), we can further approximate (\ref{residueatinftygpds}) as:
\begin{align}
  F(x, \xi, 0) & \approx-2 \sqrt{2} {C(z)}\operatorname{Res}\left[x^{-j-1}\left[\frac{2}{1+\sqrt{1-z}}\right]^j \mathcal{F}_j(\xi, t) ; \infty\right]\ , \\
  & =-C_p(z) \operatorname{Res}\left[x_p(z)^{-(j+1)} \mathcal{F}_j(\xi, t) ; \infty\right]\ ,
\end{align}
where $x_p(z)=\frac{1+\sqrt{1-z}}{2} x, C_p(z)=\left(\frac{2 \sqrt{1-z}}{1+\sqrt{1-z}}\right)^{1 / 2}$. We find that this is essentially another Mellin transformation of $\mathcal{F}_j(\xi, t)$, but transform $\mathcal{F}_j(\xi, t)$ from moment space to $x_p(z)$ space instead of the normal $x$ space. Therefore, employing the notation as (\ref{invMellinandresidue}), we have
\begin{equation}
    F(x, \xi, 0)=C_p(z)\left\{\mathcal{M}^{-1} \mathcal{F}_j(\xi, t)\right\}\left(x_p(z)\right)\ ,
\end{equation}
in the case of $\xi=0$, it reduces to the PDFs case (\ref{invMellinandresidue}). For the moment (\ref{momentrevised}), we can derive the approximation formula for GPDs when $x>\xi$
\begin{equation}
    F(x,\xi,0)=\left(\frac{2\sqrt{1-\xi^2/x^2}}{1+\sqrt{1-\xi^2/x^2}}\right)^{1/2}\left(\frac{1+\sqrt{1-\xi^2/x^2}}{1+\sqrt{1-\xi^2}}x\right)^{-\alpha}\left(1-\frac{1+\sqrt{1-\xi^2/x^2}}{1+\sqrt{1-\xi^2}}x\right)^\beta,\quad x>\xi\ ,
\end{equation}
which reduces to $x^{-\alpha}(1-x)^{\beta}$ in the forward limit of $\xi=0.$ Again, 
we caution that no polynomiality condition has been imposed on the model moments. 
In Fig. \ref{approxmbint}, we show the comparison between approximation formula and Mellin-Barnes integrals.
The approximation formula is only valid when \(x > \xi\) and \(x\) is not too close to \(\xi\). This limitation arises due to the steepest descent method used in \cite{parisAsymptoticsGaussHypergeometric2013}. When $x\sim\xi,$ the effect from the double saddle will invalidate the asymptotic expansion. Here, we only focus on analyzing the analytic properties of the Mellin-Barnes integral for \(x > \xi\), corresponding to the PDF-like region. In the DA-like region, the situation becomes more intricate, and the Mellin-Barnes integral lacks a straightforward residue interpretation.
\section{Analytic continuation of summation indices\\ and multiple Mellin-Barnes integrals}
\label{section4}
In this section, we first review the framework of double summations to calculate the amplitudes when considering the non-diagonal scale evolution operator contracting with the conformal moment or the Wilson coefficient at NLO in DVCS and DVMP. Then, we develop the technique of analytic continuation for summations indices. Finally, we convert the double summations into double Mellin-Barnes integrals to express the amplitudes, which will be beneficial for NLO calculations in future GUMP programs. This technique can be straightforwardly extended to the case of multiple summations and multiple Mellin-Barnes integrals.
\subsection{Amplitudes as double summations for DVCS and DVMP}
In the global analysis of GPDs, exclusive processes like DVCS and DVMP are crucial due to their direct correlation with the experimental quantities such as the Compton form factor and transition form factor. To achieve a universal moment parametrization as GUMP underlined, we mainly work in the conformal moment space. Schematically, we can write the amplitudes of DVCS \cite{kumerickiFittingProcedureDeeply2008} and DVMP \cite{Muller:2013jur} as
\begin{equation}
  \mathcal{A}\propto\sum_{i=q, \bar{q}, g} \sum_{j=0}^{\infty}\left[1 \mp(-1)^j\right] \xi^{-j-1} C_{j}^i(Q) \mathcal{F}_j^i(\xi, t, Q) \,
  \label{amplitudes}
\end{equation}
where \(i\) is the flavor index, $Q$ is the hard scale, \(C_j(Q)\) is the \(j\)-th conformal moment of the Wilson coefficient, and the sign \(\mp\) corresponds to vector GPDs and axial-vector GPDs, respectively. We have chosen both the renormalization and factorization scales to be \(Q\). When considering the evolution of the moment or the Wilson coefficient, there will be another summation for the conformal index. The evolved GPD moment or Wilson coefficient at scale \(Q\) can be written in terms of the evolution operator \(E_{jk}^{ii'}(\xi, Q, \mu_0)\) and the unevolved ones at scale \(\mu_0\). 

In the case of moment evolution,
\begin{equation}
  F_j^i(\xi, t, \mu) = \sum_{i' = q, \bar{q}, g} \sum_{k = 0}^j \left[\frac{1 \mp (-1)^k}{2}\right] E_{jk}^{ii'}(\xi, Q, \mu_0) \mathcal{F}_k^{i'}(\xi, t, \mu_0)\ ,
  \label{momentevo}
\end{equation}
where the evolution operator \(E_{jk}^{ii'}(\xi, Q, \mu_0)\) includes non-trivial mixing in both the flavor space and conformal moment space at NLO \cite{kumerickiFittingProcedureDeeply2008}. Inserting (\ref{momentevo}) into (\ref{amplitudes}), we obtain the double summation of conformal moment space indices $j$ and $k$. We neglect flavor mixing and focus solely on the summation in conformal moment spaces, which is the primary subject of our interest here. We take the example for vector GPDs
\begin{align}
  \mathcal{A}\propto&\sum_{j=0}^{\infty} \sum_{k=0}^{j}\left[1-(-1)^j\right] \xi^{-j-1} C_{j}^i(Q)\left[\frac{1-(-1)^k}{2}\right] E_{j k}^{i i^{\prime}}\left(\xi, Q, \mu_0\right) \mathcal{F}_k^{i^{\prime}}\left(\xi, t, \mu_0\right)\ ,\\
  \label{doublesum1}=&\frac{1}{2}\sum_{j=2}^{\infty} \left[1-(-1)^j\right] C_{j}^i(Q)\sum_{k=1}^{j-1}\left[{1-(-1)^k}\right] \xi^{-k-1}E_{j k}^{i i^{\prime}}\left(Q, \mu_0\right) \mathcal{F}_k^{i^{\prime}}\left(\xi, t, \mu_0\right)\ .
\end{align}
In the second equality, we sum \(k\) from \(1\) to \(j-1\) to avoid the pole at \(k = j\) from the Regge term of the GPD moment and $k=0$ from anomalous dimension in the NLO evolution operator \cite{kumerickiFittingProcedureDeeply2008}. Besides, we have factorized out the $\xi$ dependence in the evolution kernel, which is valid at least to the NLO,
\begin{equation}
    E_{j k}^{i i^{\prime}}\left(\xi, Q, \mu_0\right)=\xi^{j-k} E_{j k}^{i i^{\prime}}\left( Q, \mu_0\right)\ .
\end{equation}

Alternatively, we can change the order of summation and relabel the indices as:
\begin{align}
  \mathcal{A}\propto&\sum_{j=0}^{\infty} \sum_{k=j}^{\infty}\left[1-(-1)^k\right] \xi^{-k-1} C_{k}^i(Q)\left[\frac{1-(-1)^j}{2}\right] E_{k j}^{i i^{\prime}}\left(\xi, Q, \mu_0\right) \mathcal{F}_j^{i^{\prime}}\left(\xi, t, \mu_0\right), \\
  \label{doublesum2}=&\frac{1}{2}\sum_{j=1}^{\infty} \left[{1-(-1)^j}\right]\xi^{-j-1}\mathcal{F}_j^{i^{\prime}}\left(\xi, t, \mu_0\right)\sum_{k=j+1}^{\infty}\left[1-(-1)^k\right]  C_{k}^i(Q) E_{k j}^{i i^{\prime}}\left( Q, \mu_0\right)\ , 
\end{align}
which physically corresponds to the evolution of Wilson coefficient physically. In principle, these two summations should produce identical results after resummation. A proper resummation technique is needed to first resum the inner summation with variable bounds, then the outer infinite summation. For the single summation as (\ref{amplitudes}) without considering evolution in conformal moment space, it is straightforward to transform into the Mellin-Barnes integral as shown in ref.\cite{Cuic:2023mki,kumerickiFittingProcedureDeeply2008}. However, for the case of double summations, the transformation to the double Mellin-Barnes integrals remains unclear in the literature.
\subsection{Analytic continuation technique and double Mellin-Barnes integrals}
\label{finiteanalytic}
In this scenario, we cannot simply regard the bounds of summation as integers. Instead, for the first part, we must determine the analytic continuation for the summation indices in (\ref{doublesum1}) and (\ref{doublesum2}). These sums are then converted into Mellin-Barnes integrals using the Sommerfeld-Watson transformation. For instance, we need to generalize the inner summation of equation (\ref{doubleresum1}) as follows:
\begin{equation}
    F_G(j):=\sumr_{k=1}^{j-1} \left[1 - (-1)^k\right] \xi^{-k-1} E_{jk}^{ii'} \mathcal{F}_k^{i'}\ ,
\end{equation}
where $\sumr$ is an formal notation indicating the analytic continuation of summation indices, so that \(j\) can take complex values. This requires finding an analytic function whose values at integer \(j\) match the finite summation. For notation simplicity, we have abbreviated \(E_{jk}^{ii'}\) for \(E_{jk}^{ii'}\left( Q, \mu_0\right)\), \(\mathcal{F}_k^{i'}\) for \(\mathcal{F}_k^{i'}\left(\xi, t, \mu_0\right)\), and $C_j^i$ for $C_j^i(Q).$
This generalization is thoroughly studied in the monograph \cite{alabdulmohsinSummabilityCalculus2018}. That is,
\begin{align}
  &\quad\sumr_{k=1}^{j-1}  \left[1 - (-1)^k\right] \xi^{-k-1} E_{jk}^{ii'} \mathcal{F}_k^{i'}\ ,\\
  &=\sum_{k=0}^\infty\left\{\left[1+(-1)^k\right] \xi^{-k-2} E_{j,k+1}^{ii'} \mathcal{F}_{k+1}^{i'}-\left[1-(-1)^{k+j}\right] \xi^{-k-j-1} E_{j,k+j}^{ii'} \mathcal{F}_{k+j}^{i'}\right\}\ ,\\
  &=-\frac{1}{2i}\int_{c_k-i \infty}^{c_k+i \infty}dk\left\{\cot{\left(\frac{\pi k}{2}\right)} \xi^{-k-2} E_{j,k+1}^{ii'} \mathcal{F}_{k+1}^{i'}-\left[\cot{(\pi k)}-\frac{1}{\sin (\pi k)}(-1)^j\right] \xi^{-k-j-1} E_{j,k+j}^{ii'} \mathcal{F}_{k+j}^{i'}\right\}\ ,
\end{align}
analogously, for the infinite inner summation of (\ref{doublesum2}), we have
\begin{align}
  &\quad\sumr_{k=j+1}^{\infty} \left[1-(-1)^k\right] C_{k}^i E_{k j}^{i i^{\prime}}\ ,\\
  &=\sum_{k=0}^\infty \left[1+(-1)^k(-1)^j\right]C_{k+j+1}^i E_{k+j+1, j}^{i i^{\prime}}\ ,\\
  &=-\frac{1}{2i} \int_{c_k-i \infty}^{c_k+i \infty}dk\left[\cot{(\pi k)}+\frac{1}{\sin{(\pi k)}}(-1)^j\right] C_{k+j+1}^i E_{k+j+1, j}^{i i^{\prime}}\ .
\end{align}
The second part should then be straightforward, as we have extended the first summation as an ordinary analytic function. We have for (\ref{doublesum1}) the moment evolution case 
\begin{align}
  &\quad\frac{1}{2}\sum_{j=2}^{\infty} \left[1-(-1)^j\right] C_{j}^i\sum_{k=1}^{j-1} \left[{1-(-1)^k}\right]\xi^{-k-1} E_{j k}^{i i^{\prime}}\mathcal{F}_k^{i^{\prime}}\ ,\\
  &=\frac{1}{2}\sum_{j=2}^{\infty} \left[1-(-1)^j\right] C_{j}^i\sumr_{k=1}^{j-1} \left[{1-(-1)^k}\right]\xi^{-k-1} E_{j k}^{i i^{\prime}}\mathcal{F}_k^{i^{\prime}}\ ,\\
  &=-\frac{1}{4i}\sum_{j=2}^{\infty} \left[1-(-1)^j\right] C_{j}^i\int_{c_k-i \infty}^{c_k+i \infty}dk\left\{\cot{\left(\frac{\pi k}{2}\right)} \xi^{-k-2} E_{j,k+1}^{ii'} \mathcal{F}_{k+1}^{i'}-\left[\cot{(\pi k)}-\frac{1}{\sin (\pi k)}(-1)^j\right] \xi^{-k-j-1} E_{j,k+j}^{ii'} \mathcal{F}_{k+j}^{i'}\right\}\ ,\\
  &=-\frac{1}{4i}\sum_{j=2}^{\infty} \left[1-(-1)^j\right] C_{j}^i\int_{c_k-i \infty}^{c_k+i \infty}dk\left\{\cot{\left(\frac{\pi k}{2}\right)} \xi^{-k-2} E_{j,k+1}^{ii'} \mathcal{F}_{k+1}^{i'}-\left[\cot{(\pi k)}+\frac{1}{\sin (\pi k)}\right] \xi^{-k-j-1} E_{j,k+j}^{ii'} \mathcal{F}_{k+j}^{i'}\right\}\ ,\\
  &=\frac{1}{8}\int_{c_j-i \infty}^{c_j+i \infty} \mathrm{d} j\tan\left(\frac{\pi j}{2}\right) C_{j+2}^i\int_{c_k-i \infty}^{c_k+i \infty} \mathrm{d} k\cot\left(\frac{\pi k}{2}\right)\left(\xi^{-k-2} E_{j+2, k+1}^{i i^{\prime}} \mathcal{F}_{k+1}^{i^{\prime}}-\xi^{-k-j-3} E_{j+2, k+j+2}^{i i^{\prime}} \mathcal{F}_{k+j+2}^{i^{\prime}}\right)\ ,
\end{align}
similarly for (\ref{doublesum2}) the coefficient evolution case, we can derive
\begin{align}
  &\quad\frac{1}{2}\sum_{j=1}^{\infty} \left[{1-(-1)^j}\right]\xi^{-j-1}\mathcal{F}_j^{i^{\prime}}\sum_{k=j+1}^{\infty}\left[1-(-1)^k\right]  C_{k}^i E_{k j}^{i i^{\prime}}\ , \\
  &=\frac{1}{2}\sum_{j=1}^{\infty} \left[{1-(-1)^j}\right]\xi^{-j-1}\mathcal{F}_j^{i^{\prime}}\sumr_{k=j+1}^{\infty}\left[1-(-1)^k\right]  C_{k}^i E_{k j}^{i i^{\prime}}\ ,\\
  &=-\frac{1}{4i}\sum_{j=1}^{\infty} \left[{1-(-1)^j}\right]\xi^{-j-1}\mathcal{F}_j^{i^{\prime}}\int_{c_k-i \infty}^{c_k+i \infty}dk\left[\cot{(\pi k)}+\frac{1}{\sin{(\pi k)}}(-1)^j\right] C_{k+j+1}^i E_{k+j+1, j}^{i i^{\prime}}\ ,\\
  &=-\frac{1}{4i}\sum_{j=1}^{\infty} \left[{1-(-1)^j}\right]\xi^{-j-1}\mathcal{F}_j^{i^{\prime}}\int_{c_k-i \infty}^{c_k+i \infty}dk\left[\cot{(\pi k)}-\frac{1}{\sin{(\pi k)}}\right] C_{k+j+1}^i E_{k+j+1, j}^{i i^{\prime}}\ ,\\
  &=\frac{1}{4i}\sum_{j=0}^{\infty} \left[{1+(-1)^j}\right]\xi^{-j-2}\mathcal{F}_{j+1}^{i^{\prime}}\int_{c_k-i \infty}^{c_k+i \infty}dk\tan\left(\frac{\pi j}{2}\right) C_{k+j+2}^i E_{k+j+2, j+1}^{i i^{\prime}}\ ,\\
  &=\frac{1}{8}\int_{c_j-i \infty}^{c_j+i \infty} \mathrm{d} j\cot\left(\frac{\pi j}{2}\right)\xi^{-j-2}\mathcal{F}_{j+1}^{i^{\prime}}\int_{c_k-i \infty}^{c_k+i \infty} \mathrm{d} k\tan\left(\frac{\pi k}{2}\right)  C_{k+j+2}^i E_{k+j+2, j+1}^{i i^{\prime}}\ .
\end{align}
To summarize, at the end of the day,  we have for the vector GPDs
\begin{multline}
\mathcal{A}_V\propto\frac{1}{8}\int_{c_j-i \infty}^{c_j+i \infty} \mathrm{d} j\left[i+\tan\left(\frac{\pi j}{2}\right)\right] C_{j+2}^i\int_{c_k-i \infty}^{c_k+i \infty} \mathrm{d} k\cot\left(\frac{\pi k}{2}\right)\\\left(\xi^{-k-2} E_{j+2, k+1}^{i i^{\prime}} \mathcal{F}_{k+1}^{i^{\prime}}-\xi^{-k-j-3} E_{j+2, k+j+2}^{i i^{\prime}} \mathcal{F}_{k+j+2}^{i^{\prime}}\right)\ ,
\label{doubleresum1}
\end{multline}
as well as
\begin{equation}
\mathcal{A}_V\propto-\frac{1}{8}\int_{c_j-i \infty}^{c_j+i \infty} \mathrm{d} j\left[i-\cot\left(\frac{\pi j}{2}\right)\right]\xi^{-j-2}\mathcal{F}_{j+1}^{i^{\prime}}\int_{c_k-i \infty}^{c_k+i \infty} \mathrm{d} k\tan\left(\frac{\pi k}{2}\right)  C_{k+j+2}^i E_{k+j+2, j+1}^{i i^{\prime}}\ ,
\label{doubleresum2}
\end{equation}
where \(c_j\) and \(c_k\) are both chosen to be in the range \((-1,0)\) and to the right of all the poles of $\mathcal{F}_{j/k}^{i'}.$ Note that we have manually added an extra \(i\) to the \(j\) integral in accordance with \cite{Cuic:2023mki,kumerickiFittingProcedureDeeply2008} to account for the imaginary part of the amplitude. For axial-vector GPDs, the treatment is alike
\begin{multline}
  \mathcal{A}_P\propto-\frac{1}{8}\int_{c_j-i \infty}^{c_j+i \infty} \mathrm{d} j\left[i-\cot\left(\frac{\pi j}{2}\right)\right] C_{j+2}^i\int_{c_k-i \infty}^{c_k+i \infty} \mathrm{d} k\left\{\tan\left(\frac{\pi k}{2}\right)\xi^{-k-2} E_{j+2, k+1}^{i i^{\prime}} \mathcal{F}_{k+1}^{i^{\prime}}\right.\\\left.+\cot\left(\frac{\pi k}{2}\right)\xi^{-k-j-3} E_{j+2, k+j+2}^{i i^{\prime}} \mathcal{F}_{k+j+2}^{i^{\prime}}\right\}\ ,
  \label{doubleresum3}
\end{multline}
and 
  \begin{equation}
  \mathcal{A}_P\propto-\frac{1}{8}\int_{c_j-i \infty}^{c_j+i \infty} \mathrm{d} j\left[i+\tan\left(\frac{\pi j}{2}\right)\right]\xi^{-j-2}\mathcal{F}_{j+1}^{i^{\prime}}\int_{c_k-i \infty}^{c_k+i \infty} \mathrm{d} k\tan\left(\frac{\pi k}{2}\right)  C_{k+j+2}^i E_{k+j+2, j+1}^{i i^{\prime}}\ .
  \label{doubleresum4}
  \end{equation}
We have tested that, providing numerical implementation ensuring proper evaluation and reliable convergence, these two double Mellin-Barnes integrals for the evolution of either the conformal moment or the Wilson coefficient both yield the same result. These double integrals can be utilized in future GUMP programs. 

For the case of multiple summations, this approach remains available. Similar methods of analytically continuing the summation indices and applying the Sommerfeld-Watson transformation can effectively convert these multiple summations into multiple Mellin-Barnes integrals.

\section{Conclusion}

In this work, we investigate the convergence properties of the conformal moment expansion of GPDs useful for their global analysis from various constraints, elucidating the relationship between formal summation and the Mellin-Barnes integral, which had previously been suggested in earlier GUMP studies \cite{Guo:2022upw, Guo:2023ahv}. Furthermore, we examine the asymptotic properties of the Mellin-Barnes integral and derive asymptotic conditions on the conformal moment \(\mathcal{F}_j\) that had not been revealed in prior research. A limitation of this formulation is that the derived asymptotic conditions appear to contradict the polynomiality condition of conformal moments which can be remedied easily. 

Furthermore, we propose an approximate formula for the GPDs \(F(x, \xi, t)\) for \(x > \xi\). Note that \(x\) cannot be too close to \(\xi\) due to the limitations of the steepest descent method used to find the asymptotic expansion of hypergeometric functions. Identifying the uniform asymptotic expansion of this hypergeometric function would enable us to overcome this limitation, although achieving this would necessitate considerable mathematical rigor.

Finally, we introduce a method to effectively manage double summations arising from the NLO non-diagonal evolution kernel in DVCS and DVMP. By converting these summations into multiple Mellin-Barnes integrals through the analytic continuation of summations with variable bounds, we establish a mathematical framework. This technique is anticipated to be beneficial for future research within the GUMP program, as it provides a foundation for accurately calculating amplitudes at NLO, which is crucial for comprehensive global analysis.

\begin{acknowledgments}

We thank Yuxun Guo, M.~Gabriel~Santiago and Yushan Su for useful discussions. XJ is partially supported by Maryland Center for Fundamental Physics (MCFP), and HCZ acknowledges the hospitality and support of MCFP and U. Maryland for the summer intern program. 

\end{acknowledgments}

\appendix

\section{Sommerfeld-Watson transformation}
\label{Sommerfeld}
The Sommerfeld-Watson transformation is a technique for converting series into complex integrals, based on the residue theorem, proposed in \cite{sommerfeldPartialDifferentialEquations1964,DiffractionElectricWaves1918}. For our usage, we have the following formulas:
\begin{equation}
  \sum_{k=0}^{\infty}(-1)^k f_k=-\frac{1}{2 i} \int_{c-i \infty}^{c+i \infty} \frac{1}{\sin (\pi s)} f_s ds\ ,
\end{equation}
and
\begin{equation}
  \sum_{k=0}^{\infty} f_k=-\frac{1}{2 i} \int_{c-i \infty}^{c+i \infty} \cot{(\pi s)} f_s ds\ ,
\end{equation}
where \(f_s\) is the analytic continuation of \(f_k\), having no poles on the right half-plane. Here, \(c\) is chosen such that \(-1 < c < 0\) and to be to the right of all the poles of \(f_s\). Furthermore, we have
\begin{equation}
  \sum_{k=0}^{\infty}\left[1-(-1)^k\right] f_k=\frac{1}{2 i} \int_{c-i \infty}^{c+i \infty} \tan \left(\frac{\pi s} {2}\right) f_s ds\ ,
\end{equation}
and 
\begin{equation}
  \sum_{k=0}^{\infty}\left[1+(-1)^k\right] f_k=-\frac{1}{2 i} \int_{c-i \infty}^{c+i \infty} \cot \left(\frac{\pi s}{2}\right) f_sds\ .
\end{equation}

\bibliographystyle{jhep}
\bibliography{bib}

\end{document}